\documentclass[aps,pra,superscriptaddress,twocolumn]{revtex4-1}

\usepackage{graphicx,subfigure}
\usepackage{amssymb,amsmath,amsfonts,mathrsfs,bm}

\usepackage{dsfont}
\usepackage{color}

\DeclareMathOperator{\sech}{sech}
\DeclareMathOperator{\Tr}{Tr}

\newcommand{\fr}[2]{{\textstyle\frac{#1}{#2}}}

\newcommand{\bS}{{\bm S}}
\newcommand{\qA}{{\rm{A}}} 
\newcommand{\qB}{{\rm{B}}} 

\newcommand{\SA}{{\bm S}_{{\rm{A}}}}
\newcommand{\SB}{{\bm S}_{{\rm{B}}}}
\newcommand{\hSA}{\hat{\bm S}_{{\rm{A}}}}
\newcommand{\hSB}{\hat{\bm S}_{{\rm{B}}}}
\newcommand{\nA}{n_{{\rm{A}}}}
\newcommand{\nB}{n_{{\rm{B}}}}
\newcommand{\gA}{g_{{\rm{A}}}}
\newcommand{\gB}{g_{{\rm{B}}}}
\newcommand{\hA}{h_{{\rm{A}}}}
\newcommand{\hB}{h_{{\rm{B}}}}

\newcommand{\hsigA}{\hat\sigma_{{\rm{A}}}}
\newcommand{\hsigB}{\hat\sigma_{{\rm{B}}}}
\newcommand{\hbsigA}{\hat{\bm\sigma}_{{\rm{A}}}}
\newcommand{\hbsigB}{\hat{\bm\sigma}_{{\rm{B}}}}

\newcommand{\cH}{{\cal H}}
\newcommand{\cU}{{\cal U}}

\newcommand{\hilb}[1]{\mathscr{H}_{#1}}
\newcommand{\ket}[1]{|{#1}\rangle}
\newcommand{\bra}[1]{\langle{#1}|}
\newcommand{\braket}[2]{\langle{#1}|{#2}\rangle}
\newcommand{\bket}[1]{\big|{#1}\big\rangle}
\newcommand{\bbra}[1]{\big\langle{#1}\big|}
\newcommand{\bbraket}[2]{\big\langle{#1}\big|{#2}\big\rangle}

\begin{document}
	
\title{
Quantum correlations between distant qubits conveyed by large-$S$ spin chains
}

\author{Davide Nuzzi}
\affiliation{Dipartimento di Fisica, Universit\`a di Firenze,
	Via G. Sansone 1, I-50019 Sesto Fiorentino (FI), Italy}

\affiliation{INFN Sezione di Firenze, via G.Sansone 1,
	I-50019 Sesto Fiorentino (FI), Italy}

\author{Alessandro  Cuccoli}
\affiliation{Dipartimento di Fisica, Universit\`a di Firenze,
             Via G. Sansone 1, I-50019 Sesto Fiorentino (FI), Italy}

\affiliation{INFN Sezione di Firenze, via G.Sansone 1,
             I-50019 Sesto Fiorentino (FI), Italy}

\author{Ruggero Vaia}
\affiliation{Istituto dei Sistemi Complessi,
             Consiglio Nazionale delle Ricerche,
             via Madonna del Piano 10,
             I-50019 Sesto Fiorentino (FI), Italy}

\affiliation{INFN Sezione di Firenze, via G.Sansone 1,
	I-50019 Sesto Fiorentino (FI), Italy}

\author{Paola Verrucchi}
\affiliation{Istituto dei Sistemi Complessi,
             Consiglio Nazionale delle Ricerche,
             via Madonna del Piano 10,
             I-50019 Sesto Fiorentino (FI), Italy}
\affiliation{Dipartimento di Fisica, Universit\`a di Firenze,
             Via G. Sansone 1, I-50019 Sesto Fiorentino (FI), Italy}
\affiliation{INFN Sezione di Firenze, via G.Sansone 1,
             I-50019 Sesto Fiorentino (FI), Italy}
\date{\today}

\begin{abstract} 
We consider two distant spin-$\frac{1}{2}$ particles 
(or {\it qubits}) and a number of interacting objects, all with the 
same value $S\gg1$ of their respective spin, distributed on a 
one-dimensional lattice (or {\it large-$S$ spin chain}). The quantum 
states of the chain are constructed by linearly combining tensor 
products of single-spin coherent states, whose evolution is determined 
accordingly, i.e., via classical-like equations of motions. We show that 
the quantum superposition of the above product states resulting from a 
local interaction between the first qubit and one spin of the chain 
evolves so that the second qubit, after having itself interacted with 
another spin of the chain, can be entangled with the first qubit. 
Obtaining such outcome does not imply imposing constraints on the length 
of the chain or the distance between the qubits, which demonstrates the 
possibility of generating quantum correlations at a distance by means of 
a macroscopic system, as far as local interactions with just a few of 
its components are feasible. 
\end{abstract}

\pacs{03.67.Bg, 75.10.Pq, 05.45.Yv, 03.67.Lx}
% 03.67.Lx Quantum computation
% 03.67.Bg Entanglement production and manipulation
% 05.45.Yv Solitons
% 75.10.Pq Spin chain models
% 75.10.Hk Classical spin models
% 03.67.Hk Quantum communication

\maketitle

\section{Introduction}
\label{s.intro}

Whenever dealing with quantum devices one needs to accommodate 
antithetical requirements: On the one hand, microscopic objects must be 
isolated from their environment to protect the quantum behavior which is 
key to the device functioning. On the other hand, they need  
communicating with the external world in order to accomplish some useful 
task. This suggests that a hybrid scheme might be necessary in order 
to meet both requirements, where by hybrid we mean a system 
where the fragile quantum component (one or more qubits) 
is accompanied by a robust, almost classical partner, which mediates the 
dialog between each qubit and the external world without significantly 
exposing it, but still being able of conveying quantum correlations.

Specifically addressing the case of quantum operations such as 
state transfer or entanglement generation, the most promising 
proposals are typically based on 
the use of quantum channels made of interacting qubits~ 
\cite{Bose2003,CamposVenutiEtal2007,BanchiEtal2011,CampbellEtAl2011,ApollaroBCVV2012,
PaganelliEtAl2013,Kay2010,KarbachS2005,ChristandlDEL2004,YungB2005,DiFrancoPK2008,
AhmedG2015,ZwickASO2011,WangSR2011},
whose expected high performances entail a high sensitivity to 
decoherence, raising the necessity of protection from external 
disturbances. This level of protection could be alleviated if it were 
possible to exploit the more robust dynamical features of a system made 
by interacting objects with a large value of their spin angular momentum, 
$S\gg1$, possibly arranged on a one-dimensional (1D) lattice, so as to make 
up the system that we will hereafter call "large-$S$ spin chain".
Indeed, a classical analysis based on the $S\to\infty$ limit, has 
recently shown that such a spin chain can be made to evolve in a way 
such that robust signals (specifically magnetic solitons) are 
transmitted along macroscopic distances, giving rise to an overall 
dynamics that fulfills single-qubit state 
manipulation~\cite{CNVV2014a,CNVV2015a}. However,  
in order to demonstrate that a large-$S$ spin-chain can also be used 
for generating entanglement between distant qubits, a quantum treatment 
of its dynamics must be considered.

The exact quantum description of large-$S$ spin chains of sizeable 
length is usually unattainable, even numerically, due to their huge Hilbert 
space and the specific algebra obeyed by the spin operators: therefore,
{\em ad hoc} methods must be devised to deal with 
such a problem. Generalized Coherent States (GCS)~\cite{Perelomov1986, 
ZhangFG1990} provide a powerful tool for describing the dynamics of 
quantum systems in this context~\cite{CalvaniCGV2013,BalakrishnanB1989,DiosiS1997,HelmSRW2011,Combescure1992}, as 
they keep a clear correspondence between the quantum picture and the 
increasingly classical behavior observed when the system quanticity 
parameter (e.g., $1/S$ for spins) tends to zero~\cite{Lieb73,Yaffe82}.

\begin{figure}
	\centering
	\includegraphics[width=0.45\textwidth]{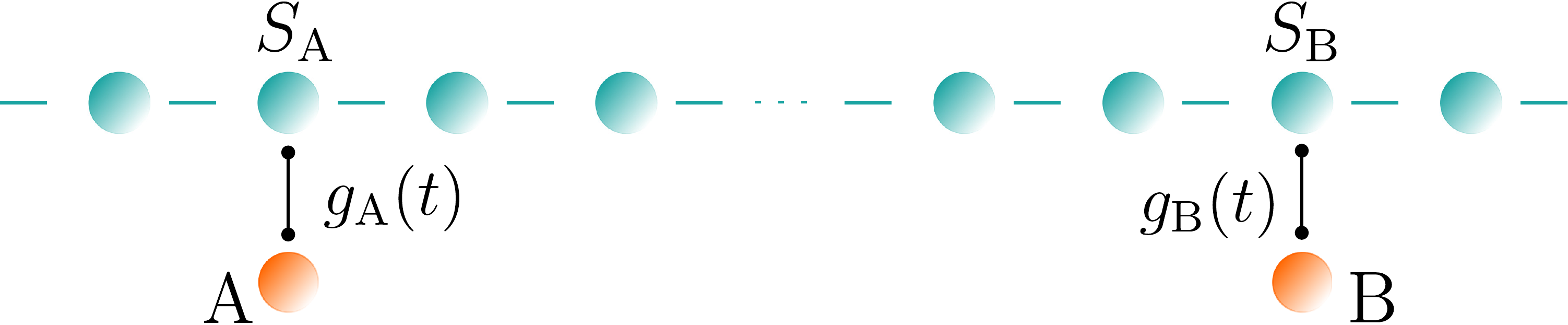}
	\caption{
	Schematic representation of the system: 
the connected blue spheres represent the spins of a Heisenberg chain, 
while the orange ones are the two qubits.}
	\label{f.schema}
\end{figure}

Aim of this work is to illustrate the possibility of generating 
entanglement between two qubits separated by a macroscopic distance by 
means of their interaction with localized components of a large-$S$ spin 
chain. This is made possible by treating such a hybrid system within an 
approximate description, based on the properties of GCS for the 
large-$S$ spin chain, that retains enough of the spin-chain quantum 
nature to account for quantum correlations. 
Specifically, we choose an isotropic Heisenberg chain, referred to as 
$\Gamma$ henceforth, composed by elementary objects with spin quantum 
number $S$ larger than $1/2$, and two external qubits $\qA$ and $\qB$ 
interacting with two spins $\SA$ and $\SB$ of the 
chain, as shown in Fig.~\ref{f.schema}.
Starting from a factorized state of the chain and the two qubits,  we 
study whether the 
entanglement locally created by the interaction between $\qA$ and $\SA$ 
can propagate along the chain up to $\SB$ and be finally transferred to 
$\qB$, the net result being the generation of entanglement between $\SA$ 
and $\SB$ .

The initial state of $\Gamma$ is taken as a tensor product of 
single-spin coherent states (SCS), which allows us to establish a one-to-one 
correspondence between the configurations of a classical spin chain and 
the quantum states of $\Gamma$. Indeed, if 
$\Gamma$ sustains the propagation of Heisenberg 
solitons~\cite{TjonW1977, Takhtajan1977} (i.e., well-localized, stable, 
pulse-shaped excitations), their propagation can 
trigger the interaction with the two qubits and convey the quantum 
correlations between them.

In Sec.~\ref{s.model} we describe the model for the overall system 
and specify the interactions between its components.
The system dynamics is then divided into 
three different stages, which are considered in Secs.~\ref{s.1st}, 
\ref{s.2nd}, and~\ref{s.3rd}. Role of Sec.~\ref{s.interlude} is that of 
providing a formal derivation of the GCS for the large-$S$ 
Heisenberg chain, 
while Sec.~\ref{s.conclusions} 
is devoted to the discussion of our numerical results and the
concluding remarks.

\section{Model set-up}
\label{s.model}

The Hamiltonian describing the overall system is taken of the form:
\begin{equation}
 \cH=\cH_{\qA{,}\SA} + \cH_{\Gamma}+\cH_{\qB{,}\SB},
\label{e.ham_full1}
\end{equation}
where  
\begin{equation}
 \cH_{\qA{,}\SA} = \gA\,\hSA{\cdot}\,\hbsigA + \hA \hsigA^z ~,
\label{e.ham_q1s1}
\end{equation}
and similarly for $\cH_{\qB{,}\SB}$, with $\qA\leftrightarrow\qB$; 
$\hbsigA$ and $\hbsigB$ are the Pauli operators of the qubits, whose 
interaction with $\SA$ and $\SB$ is ruled by the 
coupling constants $\gA$ and $\gB$, while $\hA$ and $\hB$ are uniform 
magnetic fields possibly applied to the qubits only.
The Hamiltonian of the chain 
\begin{equation}
\cH_\Gamma=-J\sum_n\hat{\bm S}_n{\cdot}\hat{\bm S}_{n+1} 
         - \gamma H \sum_n \hat{S}_n^z  \label{e.hchain} 
\end{equation} 
embodies a nearest-neighbor isotropic ferromagnetic interaction, whose 
strength is given by the exchange constant $J>0$, and that with an 
external field ${\bm H}$, $\gamma$ being the gyromagnetic ratio; amongst 
the possible solutions of the equations of 
motions 
defined by ${\cal H}_\Gamma$ in the $S\to\infty$ and continuum limit, 
for $H\neq0$ there are the so-called Heisenberg solitons \cite{TjonW1977}; 
in Appendix~\ref{a.Hsolitons} we briefly recall the
properties of such solutions that are relevant in this work.

We further
assume that the qubit-chain couplings $\gA$ and $\gB$ 
depend on time, and are switched on and off according to
\begin{equation}
\begin{aligned} 
 & \gA(t) = g \, \vartheta(t{-}t_0) \, \vartheta(t_1{-}t)~,\\
 & \gB(t) = g \, \vartheta(t{-}t_2) \, \vartheta(t_3{-}t)~,
\end{aligned}
\label{e.intconst}
\end{equation}
with $\vartheta(t)$ the Heaviside function, and $g$  the 
interaction strength. By Eqs.~\eqref{e.intconst} the overall evolution 
is decomposed into three stages, and $g$ sets the time-scale for the 
first and third ones.
\begin{figure}
	\centering
	\includegraphics[width=0.45\textwidth]{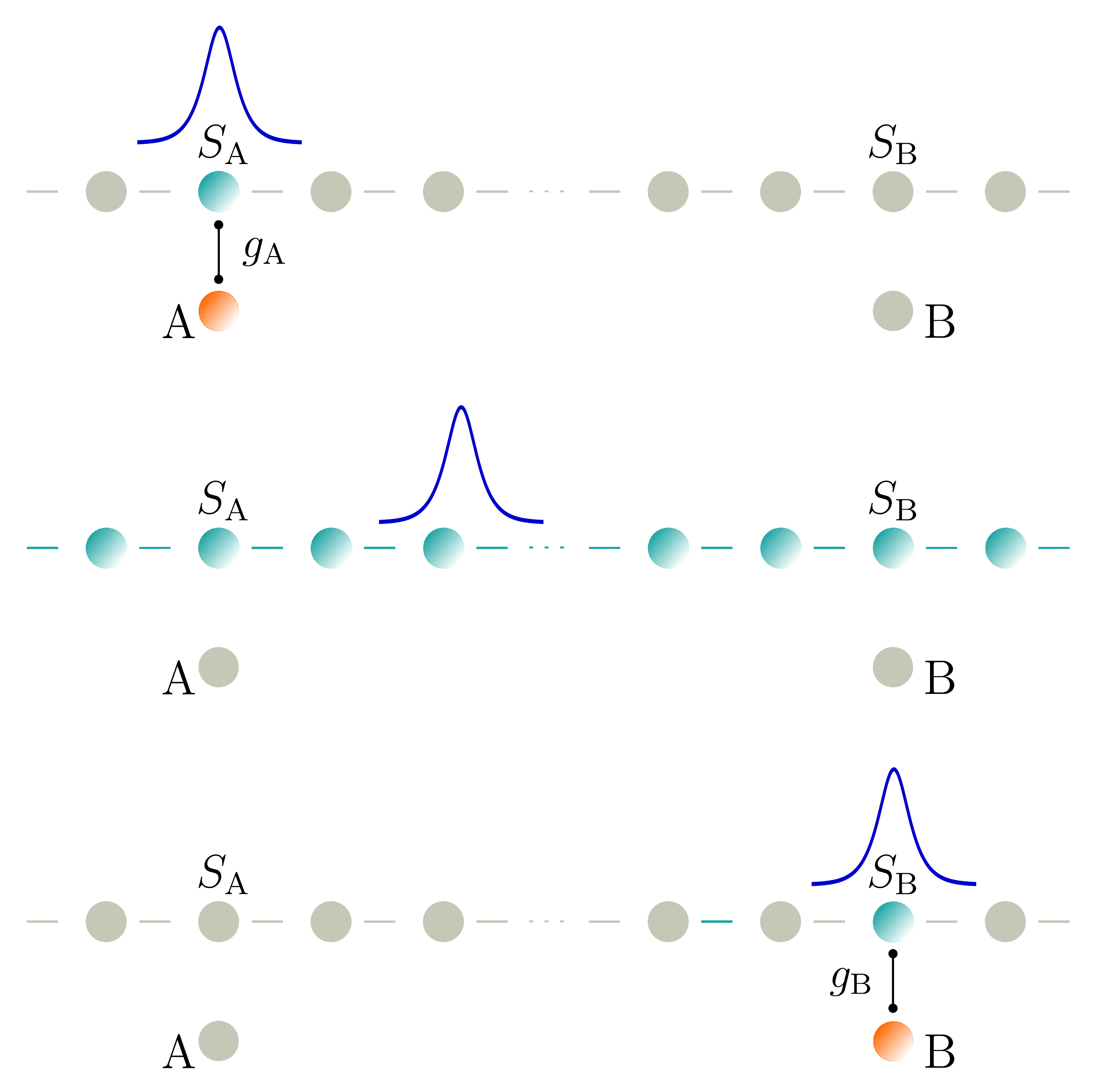}
	\caption{Schematic representation of the three stages of the system dynamics (1), (2) and (3) from top to bottom, respectively (see text).}
	\label{f.intconst}
\end{figure} 
In more detail, assuming an initial factorized state, 
the system evolution from $t_0$ to $t_3$ 
(see Fig.~\ref{f.intconst} for a graphic representation) is described as 
follows:
\begin{itemize}
 \item[(1)] {$ [t_0,t_1]$}: $\gA(t)=g$ and $\gB(t)=0$. Starting from the 
factorized initial state the evolution of $(\qA,\SA)$ is determined. 
During this stage $\Gamma\setminus\SA$ is frozen, i.e., the spins 
$\{\bS_n\}$ except $\SA$~\cite{note1} do not evolve. This leads to an 
entangled state of $\qA$ and $\SA$.
 \item[(2)] {$ [t_1,t_2]$}: $\gA(t)=\gB(t)=0$. The large-$S$ evolution 
of $\Gamma$ results in an entangled state of $\qA$ with the 
entire chain.
 \item[(3)] {$ [t_2,t_3]$}: $\gB(t)=g$ and $\gA(t)=0$. The relevant 
evolution only concerns the $(\qB,\SB)$ pair, and $\Gamma\setminus\SB$ 
is frozen. The eventual 
result is an entangled  
state of the whole system, with a finite concurrence between $\qA$ and 
$\qB$.
\end{itemize} 
As for the dynamics of the system for $t<t_0$, the qubits and 
the spins in a portion of $\Gamma$ having $\SA$ and $\SB$ well within 
its bulk, stay all aligned along the field direction; meanwhile, a 
Heisenberg soliton travels from the left towards the above chain 
portion, so as to reach $\SA$ at $t=t_0$.

\section{First stage: evolution of $(A,\SA)$}
\label{s.1st}

In the first stage we observe the evolution of the qubit $\qA$ 
interacting 
with the spin $\SA$, while all the other spins of $\Gamma$ are frozen. 
The initial state at $t=t_0$ is assumed to be
\begin{equation}
 \ket{\Psi(t_0)} = \ket{A}\otimes\left[ 
 \, \bigotimes_{n} \bket{\Omega_n(t_0)} \right]\otimes \ket{B}  ~,
\label{e.sys_iniconf}
\end{equation}
where $\ket{A}$ and $\ket{B}$ are the qubit states.
The state of $\Gamma$, in brackets in Eq.~\eqref{e.sys_iniconf}, is a 
tensor product of single spin states, which are 
chosen as SCS~\cite{Radcliffe1971}. The SCS 
form an overcomplete set on each Hilbert space $\hilb{S_n}$  
and they are in one-to-one correspondence with 
the configurations of a classical spin (namely a fixed-length vector),
which implies that they can be parametrized by polar angles (see Appendix~\ref{a.spinCS}).
The main reason for the above choice of the chain initial state
is that tensor products of SCS provide the GCS for the large-$S$ Heisenberg 
chain, as shown in the next section. In particular, the SCS that we will 
use in Eq.~\eqref{e.sys_iniconf} are those defined
by the polar angles corresponding to a Heisenberg-soliton shape,
as described by Eq.~(\ref{e.twsol})
with $x=nd$. We further enforce the condition $n_Ad=vt_0$, 
where ${\bm S}_{n_A}=\SA$, $d$ is the lattice spacing,  and $v$ is the soliton 
velocity, so that the traveling soliton be centered at $n_A$ for
$t=t_0$.

The evolution of the system during the time interval $[t_0,t_1]$ is 
described by 
\begin{equation}
 \cU^{(1)}(t) = \cU_{\qA{,}\SA}(t) \otimes \mathds{1}_{\Gamma\setminus\SA}
    \otimes \cU_{\qB}(t) ~,
\label{e.1st_evo_full}
\end{equation}
where
\begin{equation}
 \cU_{\qA{,}\SA}(t) = \exp(-i\, \cH_{\qA{,}\SA}\,t) ~,
\label{e.1st_evo}
\end{equation}
is the propagator for the subsystem ($\qA$,$\SA$), with $\cH_{\qA{,}\SA}$ 
as in Eq.~(\ref{e.ham_q1s1}),
while 
\begin{equation}
 \cU_{\qB}(t) = \exp(-i\, \hB\hsigB^z\,\,t)\,
\label{e.uq2}
\end{equation}
accounts for the effect on $\qB$ of the local field $\hB$.

\begin{figure}
	\centering
	\includegraphics[width=0.45\textwidth]{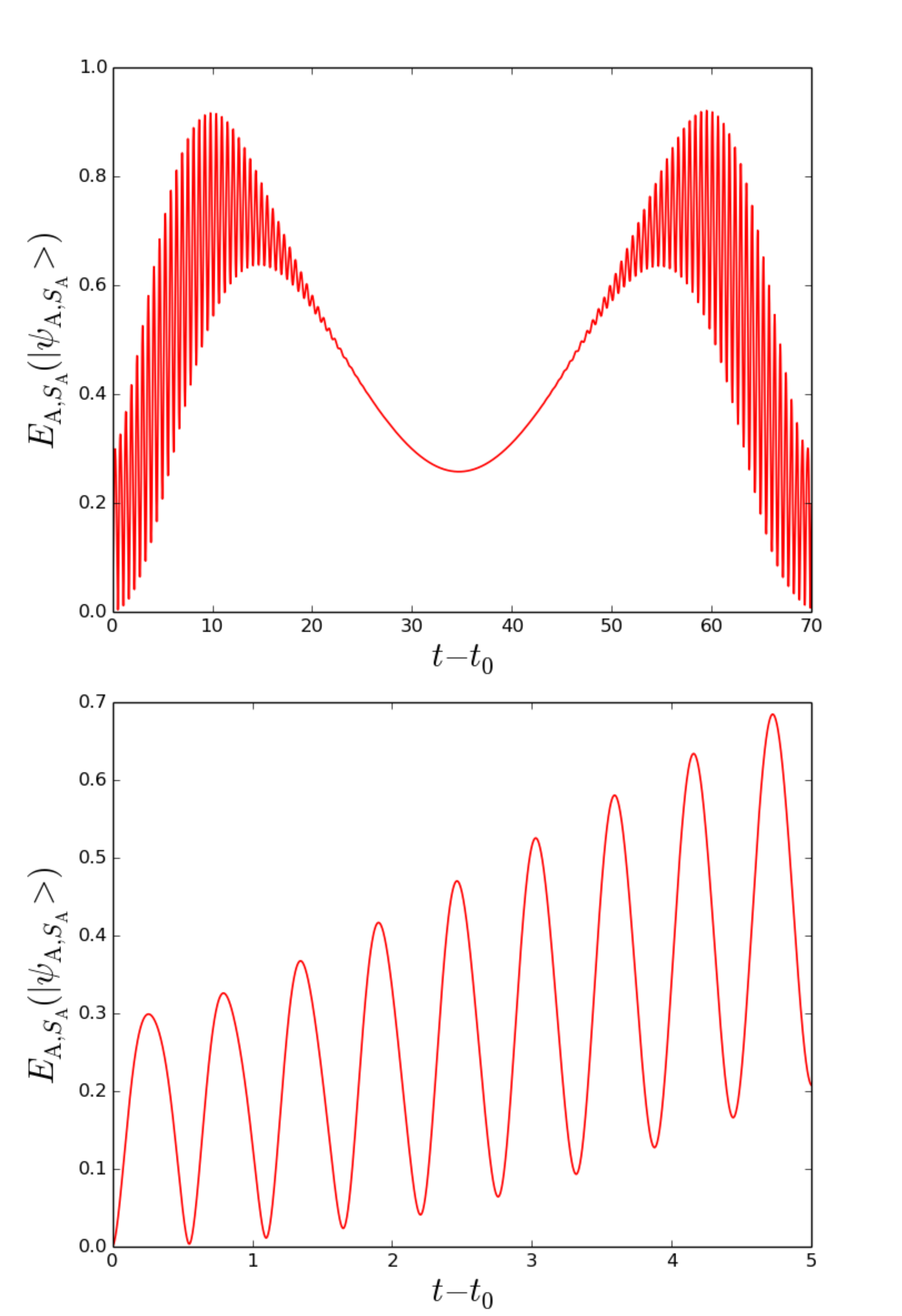}
	\caption{$E_{{\qA,\SA}}(t)$ for $t\in[t_0, t_1]$, $S = 5$, $\gA = 1$, $\hA = 0.25$. The $\qA$ initial state is $\ket1$ while the chain is initially in the state corresponding to a propagating Heisenberg soliton (see text) with $\lambda_\beta=10$ and $\beta=\pi/4$ centered in $n_A$ (specifically meaning $\Omega^0_{\nA} = \{\theta^0_{\nA} = \pi/2, \varphi^0_{\nA} = 0\} $). The lower panel shows a zoom of the small-time part of the upper one.}
	\label{f.Eq1vstime}
\end{figure}

As far as $\Gamma\setminus{\SA}$ and $\qB$ are concerned, the action of 
$\cU^{(1)}(t{-}t_0)$ on the state~\eqref{e.sys_iniconf} is trivial; 
however, the subsystem $(\qA,\SA)$ can evolve into an entangled state, 
as shown in Fig.~\ref{f.Eq1vstime}, where the Von Neumann entropy 
$E_{{\qA,\SA}}(\bket{\psi_{\qA,\SA}(t)})$ of $\qA$ is shown 
 as a function of time~\cite{note2}, for one initial state of $\SA$ and 
$\qA$ and given values of the relevant parameters
(times and lengths in figures are in reduced units of $(JS)^{-1}$ and 
$d$, respectively).

As we aim at generating entanglement between $\qA$ and $\qB$ via 
$\Gamma$, we will choose $t_1$ so as to maximize the numerically evaluated 
entanglement between $\qA$ and $\SA$ at the end of the first dynamical stage.
We thus ensure that the initial separable state of $(\qA,\SA)$ defined 
by Eq.~\eqref{e.sys_iniconf},
\begin{equation}
 \bket{\psi_{\qA,\SA}(t_0)}= \ket{A} \otimes \ket{\Omega_{\nA}(t_0)} ~,
\label{e.inistate_S1q1}
\end{equation}
will evolve~\cite{note3} into an entangled state  
\begin{equation}
 \bket{\psi_{\qA,\SA}(t)} 
  = \sum_{\sigma{m}} c_{\sigma{m}}(t) \ket{\sigma}\otimes\ket{m} ~,
\label{e.S1q1_ent_state}
\end{equation}
where $\{\ket{\sigma}\}$ and $\{\ket{m}\}$ are orthonormal basis for $\hilb{A}$ and $\hilb{\SA}$, respectively.
By the completeness relation~\eqref{e.CScompl}, the state at time $t_1$ can be written as
\begin{equation}
 \bket{\psi_{\qA{,}\SA}(t_1) } = (2S{+}1) \sum_{\sigma}
  \int \frac{d\Omega}{4\pi}~f_\sigma^\Omega~\ket{\sigma}\otimes\ket{\Omega} ~,
\label{e.S1q1_coher_t1}
\end{equation}
with
\begin{equation}
 f_\sigma^\Omega = \sum_{m} c_{\sigma{m}}(t_1)~ \braket\Omega{m}~,
\label{e.fsigmaOmega} 
\end{equation}
and the overlap $\braket\Omega{m}$ as in Eq.~\eqref{e.CS_mO}.

The evolved state of the overall system at the end of the first stage in the density-operator formalism reads
\begin{eqnarray}
 && \bket{\Psi (t_1)} \bbra{\Psi (t_1)} = \bket{\psi_{\qA,\SA}(t_1)} \bbra{\psi_{\qA,\SA}(t_1)} \otimes
\notag\\ 
 &&~~~~~~~ \bigg[\bigotimes_{n \neq \nA} \bket{\Omega_n (t_0)} \bbra{\Omega_n (t_0)} \bigg] \otimes \bket{B(t_1)}\bbra{B(t_1)}\, ,
\label{e.t1_state}
\end{eqnarray}
where $\bket{B(t_1)}=\cU_{\qB}(t_1 - t_0) \ket{B}$, and the term in brackets is the state of $\Gamma\setminus\SA$, that is left unchanged by the first-stage dynamics.

%*********************************************************
%               Formal Interlude
%*********************************************************
\section{Coherent states of the large-$S$ spin chain}
\label{s.interlude}

This section contains a formal derivation of GCS for the Heisenberg spin 
chain with $S\gg{1}$. It will be shown that in the large-$S$ limit the 
GCS become a tensor product of SCS, as defined in 
Eq.~\eqref{e.coher_def}, thus leading to the approximate evolution that 
is described in the next section.

The construction of the GCS~\cite{ZhangFG1990} for a quantum system of 
Hamiltonian $\cal{H}$ starts from writing
\begin{equation}
 {\cal H} = \sum_i b_i \hat T_i + {\rm h.c.} \,,
\label{e.general_ham}
\end{equation} 
so as to identify the Lie algebra spanned by the operators $\big\{\hat 
T_i\big\}$; the transformation group obtained by exponentiating the 
elements of such algebra is the so-called {dynamical group} (DG), 
i.e., the unitary group ruling the dynamics of the system.

Keeping in mind that we aim at considering a large-$S$ spin chain, we 
recast ${\cal H}_\Gamma$ in a form that fits to the purpose. First, since 
the energy has to stay finite, we notice that the exchange constant and 
the gyromagnetic ratio must scale with $S$ so as to guarantee that 
$J_{\rm{c}}\equiv JS^2$ and $\gamma_{\rm{c}}\equiv \gamma{S}$ have fixed, finite, 
values. We then define the operators
\begin{equation}
 \hat{a}_n \equiv \frac{\hat S^x_n+i\hat S^y_n}{\sqrt{2}\,S}
~,~~~~
 \hat{z}_n \equiv \frac{\hat S^z_n}{S}
\end{equation}
that satisfy $[\hat{a}_n,\hat{a}^\dagger_n]=S^{-1}\hat{z}_n$ and 
$[\hat{z}_n,\hat{a}_n]=S^{-1}\hat{a}_n$,
in terms of which it is
\begin{equation}
 \frac{{\cal H}_\Gamma}{J_{\rm{c}}}=- \sum_{n=1}^{N} \big( 
 \hat{a}^\dagger_n\hat{a}_{n+1}+\hat{a}^\dagger_{n+1}\hat{a}_n
 +\hat{z}_n\hat{z}_{n+1}+h\hat{z}_n\big) ,
\label{e.hxi_pm}
\end{equation}
where $h=\gamma_{\rm{c}}H/J_{\rm{c}}$, and, for the sake of simplicity, 
periodic boundary conditions are assumed. 
Now we must find a set of operators that contains the $6N$ operators
\begin{equation}
 \big\{\hat{a}^\dagger_n,\,\hat{a}_n,\,\hat{z}_n,\,
 \hat{a}^\dagger_n\hat{a}_{n\pm1},\,\hat{z}_n\hat{z}_{n+1}\big\}~
\label{e.minimal}
\end{equation}
and is closed with respect to commutation; 
the `first generation' of commutators, namely those between the above operators, yields $4N$ new bilinear operators, namely
\begin{equation}
 \Big\{\frac{\hat{a}^\dagger_n\hat{z}_{n\pm1}}S,\,
       \frac{\hat{a}_n\hat{z}_{n\pm1}}S\Big\},
\end{equation}
as well as $10N$ new trilinear operators
\begin{equation}
 \Big\{ \frac{\hat{z}_n\hat{a}^\dagger_r\hat{a}_s}S \Big\}~,
\label{e.trilinear}
\end{equation}
where $(n,r,s)$ is either a permutation of three consecutive numbers, or $(r,s)=(n,n\pm1)$, or $(r,s)=(n\pm1,n)$.

It is clear that the exact Lie algebra won't have a finite number of 
generators, since subsequent generations of order $k$ give rise to new 
independent operators with prefactor $S^{-k}$, such as 
$S^{-2}\hat{a}_n\hat{a}_{n\pm1}$. 
%from the second generation. 
This is the reason why an exact construction of the GCS for the Heisenberg chain is not possible. However, for large $S$ 
one can disregard higher generations (i.e., approximate 
$S^{-2}\simeq{0}$) and close the Lie algebra with the above operators 
(\ref{e.minimal}-\ref{e.trilinear}).

Once the Lie Algebra that generate the DG is determined, 
the GCS are obtained by the action of displacement operators 
on an arbitrary reference state; given the physical problem we are 
dealing with, this can be chosen as the ground state of 
the Hamiltonian~\eqref{e.hchain}, i.e.,
\begin{equation}
 \bket{\Lambda} \, \equiv \, \bigotimes_n \bket{ m_n{=}S }_n \,,
 \label{e.ref_state_def}
\end{equation}    
where $\hat{S}_n^z\,\ket{m_n}_n=m_n\,\ket{m_n}_n$, so that $\hat{z}_n\,\ket{\Lambda}=\,\ket{\Lambda}$.

The displacement operators are the elements of the {left coset} of
the DG with respect to the so-called {stability subgroup}, which is 
the maximal subgroup of the DG that leaves the reference state unchanged up to a constant phase-factor.
In our case the stability subgroup is generated by 
\begin{equation}
 \Big\{ \hat{z}_n,~\hat{z}_n\hat{z}_{n+1},~ \hat{a}^\dagger_n\hat{a}_{n\pm1}, \,  \frac{\hat{z}_n\hat{a}^\dagger_r\hat{a}_s}S \Big\} \,,
\label{e.stab_sg_gen}
\end{equation}
since $\bket\Lambda$ either is an eigenstate of these operators, or is 
annihilated by them.
By definition, the left-coset representatives are given by those 
elements $\tilde{u}$ providing a unique 
decomposition of any $u\in$ DG in the form
\begin{equation}
 u = \tilde{u}\, u' \,,
\end{equation}
where $u'$ belongs to the stability subgroup. Within the large-$S$ 
approximation, it appears that the general representative of the left 
coset of the stability subgroup is given by
\begin{equation}
 \tilde{u} = \exp\Big[\, \sum_{n=1}^N \Big(
 \eta_n + \zeta^+_n\frac{\hat{z}_{n+1}}S + 
\zeta^-_n\frac{\hat{z}_{n-1}}S \Big) \hat{a}^\dagger_n - {\rm h.c.}\Big]~,
\label{e.displ_op}
\end{equation}
where $\eta \equiv (\eta_1,\ldots,\eta_N)$, $\zeta^\pm\equiv 
(\zeta^\pm_1,\ldots,\zeta^\pm_N)$ are complex vectors. Since the
operators in the exponent of the above expression commute in
the large-$S$ approximation, one can write the displacement 
operator as a product of exponentials, and recast Eq.~\eqref{e.displ_op} 
as
\begin{equation}
 \tilde{u} = \bigotimes_{n=1}^N \exp\Big[\Big(\eta_n  + \zeta^+_n\frac{\hat{z}_{n+1}}S 
    + \zeta^-_n\frac{\hat{z}_{n-1}}S \Big) \hat{a}^\dagger_n - {\rm h.c.}\Big]~.
\label{e.displop}
\end{equation} 

By applying this operator to the chosen reference state 
(\ref{e.ref_state_def}) one obtains
\begin{equation}
\tilde{u}\,\bket{\Lambda}
=\bigotimes_n 
\Big[ 
e^{\xi_n\hat{a}^\dagger_n-\xi^*_n\hat{a}_n }\,\ket{m_n{=}S}_n 
\Big] ~,
\label{e.chain_GCS}
\end{equation}
with $\xi_n=\eta_n{+}(\zeta_n^+{+}\zeta_n^-)/S$.

Setting $\xi_n=(\sqrt{2}S)^{-1}e^{i \varphi}(\theta/2)$, a one-to-one 
correspondence is established between the states that make the tensor 
product in Eq.~\eqref{e.chain_GCS} and the SCS defined in
Eq.~\eqref{e.coher_def}, after recognition of the parameters $\theta$ 
and $\varphi$ as the polar angles entering the latter. 
This correspondence implies that the GCS for the spin chain in 
the large-$S$ limit, hereafter indicated by $\ket{\Omega_\Gamma}$, 
are a tensor product of SCS, each relative to one 
spin of the chain, i.e.
\begin{equation}
\ket{\Omega_\Gamma}\equiv\tilde{u}\ket{\Lambda}
=\bigotimes_n \ket{ \Omega_n };
\label{e.prod.coh.states}
\end{equation}
this result, together with the observation that the dynamical properties of 
one-dimensional magnetic systems with large $S$ are well represented by 
classical equations of motion (EoM), leads to describe the chain 
evolution as in the next section.

%*************************************************************************************

\section{Second stage: evolution of the chain}
\label{s.2nd}

When the second stage begins, at $t=t_1$, the interaction $\gA(t)$ with 
$\qA$ is quenched and the overall propagator for 
$t\in[t_1,t_2]$ can be split as
\begin{equation}
 \cU^{(2)}(t)= \cU_{\qA}(t) \otimes \cU_\Gamma(t) \otimes
                        \cU_{\qB}(t)\,,
\label{e.2nd_evo_full}
\end{equation}
where $\cU_{\qA}(t)$ is the operator on $\qA$ analogous to that in 
Eq.~\eqref{e.uq2}, while $\cU_\Gamma(t)$ is the chain propagator. 

After the results of the previous section, we consistently take that 
the dynamics of each $\ket{ \Omega_n}$ be given by the solution of the 
classical-like EoM for the chain, meaning that any initial state 
$\ket{\Omega^0_\Gamma}=\bigotimes_n\ket{\Omega^0_n}$
evolves following the dynamics of the associated classical 
configuration, $\{\bS_n(t)=S\,{\bm s}_n(t)\}$ as from Eq.~\eqref{e.CSaveS},
with ${\bm s}_n(t)$ solving the classical EoM, Eqs.~\eqref{e.classical_EoM}, 
i.e.
\begin{equation}
%\begin{aligned}
\begin{array}{@{}cl@{\quad}@{}c@{\quad}@{}c}
 \big\{ \Omega^0_n \big\}
 &\xrightarrow[]{\text{~classical~EoM~(A2)~}}
 & \big\{ \Omega_n \big( t; \{\Omega^0_n\}\big)\big\} ~, ~~
\\[2mm]
 \bigotimes_n \ket{\Omega^0_n}
 &\xrightarrow[]{\text{~~~~~~~large-}S~~~~~~~}
 &\bigotimes_n \bket{ \Omega_n \big(t,\{\Omega^0_n\}\big)} ~.
\end{array}
%\end{aligned}
\label{e.semi-class_chainevo}
\end{equation}

This prescription provides a dynamics that reproduces the correct 
evolution of the spin expectation values in the classical limit, and 
still maintains the quantum character of $\Gamma$, allowing the 
entanglement between $\qA$ and $\SA$, generated during the first stage 
of the scheme, to be transferred via the spins of the chain.
In fact, if one starts from a pure state of $\Gamma$ which 
is factorized in the SCS basis, the above evolution cannot
transfer quantum correlations, being based on the dynamics of 
separable SCS. However, the state of $\Gamma$ when the second stage 
begins is not pure, due to $\SA$ being entangled with $A$, as 
implied by Eq.~\eqref{e.S1q1_coher_t1}.
Explicitly, once applied to the initial state Eq.~\eqref{e.t1_state}, i.e. to
\begin{equation}
 \bket{\psi_{\qA,\Gamma}(t_1)}=\bket{\psi_{\qA,\SA}(t_1)} 
 \bigotimes_{n \neq \nA} \ket{\Omega_n (t_0)}~,
\end{equation}
with $\bket{\psi_{\qA,\SA}(t_1)}$ as in Eq.(\ref{e.S1q1_coher_t1}), 
the above prescription \eqref{e.semi-class_chainevo}
leads, during the second stage, to the projector
\begin{equation}
 \bket{\Psi(t)}\!\bbra{\Psi(t)}
 =\bket{\psi_{\qA,\Gamma}(t)} \! \bbra{\psi_{\qA,\Gamma}(t)} \otimes \bket{B(t)}\!\bbra{B(t)} ~,
\label{e.t2_sys_state}
\end{equation}
with $\ket{B(t)}=\cU_{{\qB}}(t)\ket{B}$ and
\begin{equation}
 \bket{\psi_{\qA,\Gamma}(t)}= {\cal{A}}\sum_\sigma\int d\Omega~f_{\sigma}^\Omega
 \bket{\sigma(t)}\bigotimes_{n}\bket{\Omega_n(t,\Omega)}\,,
\label{e.q1chain_t2state}
\end{equation}
where $\ket{\sigma(t)}=\cU_{\qA}(t-t_1)\ket\sigma$, ${\cal{A}}$ is a normalization coefficient, and we have 
dropped the unimportant dependence of $\Omega_n(t)$ on all the $\{\Omega^0_n \equiv \Omega_n (t_0) \}$ with $n\neq n_A$, retaining 
only the meaningful dependence on $\Omega\equiv\Omega^0_{n_A}$. 

In order to identify when it is worth starting the third stage of the dynamical process, i.e. what is the best 
choice for $t_2$ as far as the further entanglement generation between A and B is concerned, we consider what 
follows. Given the Hamiltonian \eqref{e.ham_full1}, the qubit $\qB$ can become entangled with other components of 
the system, including $\qA$, exclusively via the interaction with $\SB$: entanglement generation between the two qubits
can hence occur only if $\SB$ is entangled with $(\qA,\Gamma\setminus\SB)$ at $t=t_2$, and we expect its effectiveness to 
be higher if $t_2$ is such to guarantee a significant entanglement between $\qA$ and $\SB$ at the beginning of the 
third stage. Establishing when this is the case implies determining the time-dependence of the 
Von Neumann entropy $E_{{\bS_n}}$ of any spin $\bS_n$  of the chain, that quantifies the 
entanglement between $\bS_n$ and $(\qA,\Gamma\setminus\bS_n)$. This entropy reads
\begin{equation}
E_{{\bS_n}}  = - \Tr_{_{\bS_n}} \rho_{_{\bS_n}} \log_{2S{+}1} \rho_{_{\bS_n}} ~,
\end{equation}  
where
\begin{equation}
\rho_{_{\bS_n}} = \Tr_{(\qA{,}\Gamma)\setminus\bS_n} 
\bket{\psi_{\qA,\Gamma}(t)} \bbra{\psi_{\qA,\Gamma}(t)} ~.
\end{equation}
Noticing that
\begin{equation}
\Tr_{_{\bS_l}} \bket{\Omega_l(t,\Omega)}\bra{\Omega_l(t,\Omega')}
= \bbraket{\Omega_l(t,\Omega')}{\Omega_l(t,\Omega)} ~,
\end{equation} 
we find, setting $f_{\sigma\sigma'}^{\Omega\Omega'}\equiv f_{\sigma}^{\Omega}f_{\sigma'}^{\Omega'^{\displaystyle *}}$,
\begin{eqnarray}
&& \rho_{_{\bS_n}}(t) = {\cal A}^2 \Tr_{_\qA}\sum_{\sigma\sigma'} \int d\Omega d\Omega'\,
 f_{\sigma\sigma'}^{\Omega\Omega'} \,\bket{\sigma (t)}\bbra{\sigma' (t)}
\notag\\
&& ~~~~~
\times \bigg[ \prod_{l \neq n}
\bbraket{\Omega_l(t, \Omega')}{\Omega_l(t, \Omega)} \bigg]
\notag\\
&& ~~~~~
\otimes \ket{\Omega_{n}(t,\Omega)}\bra{\Omega_{n}(t,\Omega')}~.
\label{e.t2_q1q2S2state}
\end{eqnarray}

\begin{figure}
	\includegraphics[width=0.45\textwidth]{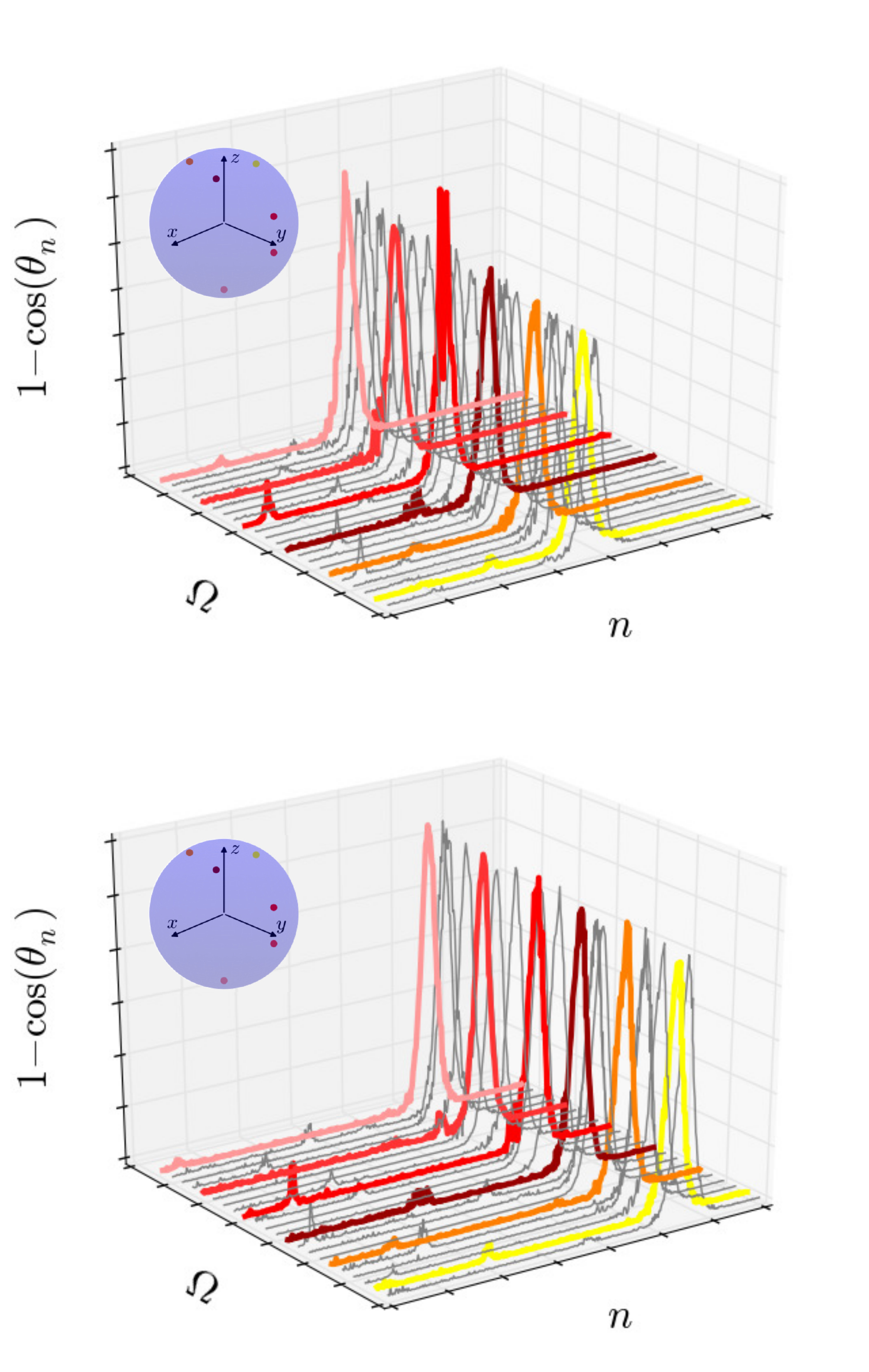}
	\caption {Configurations $\Omega_n(t,\Omega)$ 
for a starting soliton with $\lambda_\beta=10$, visualized via
$(1-\cos\theta_n(t,\Omega))$ as functions of $n$ and 
$\Omega$, after an integration time 
$t-t_1$ equal to 400 (800) in the upper (lower) panel. 
Coloured curves are for $\Omega$'s that define correspondingly
coloured points on the small sphere in the upper-left corner.}
	\label{f.3dconf}
\end{figure}
\begin{figure}
	\includegraphics[width=0.45\textwidth]{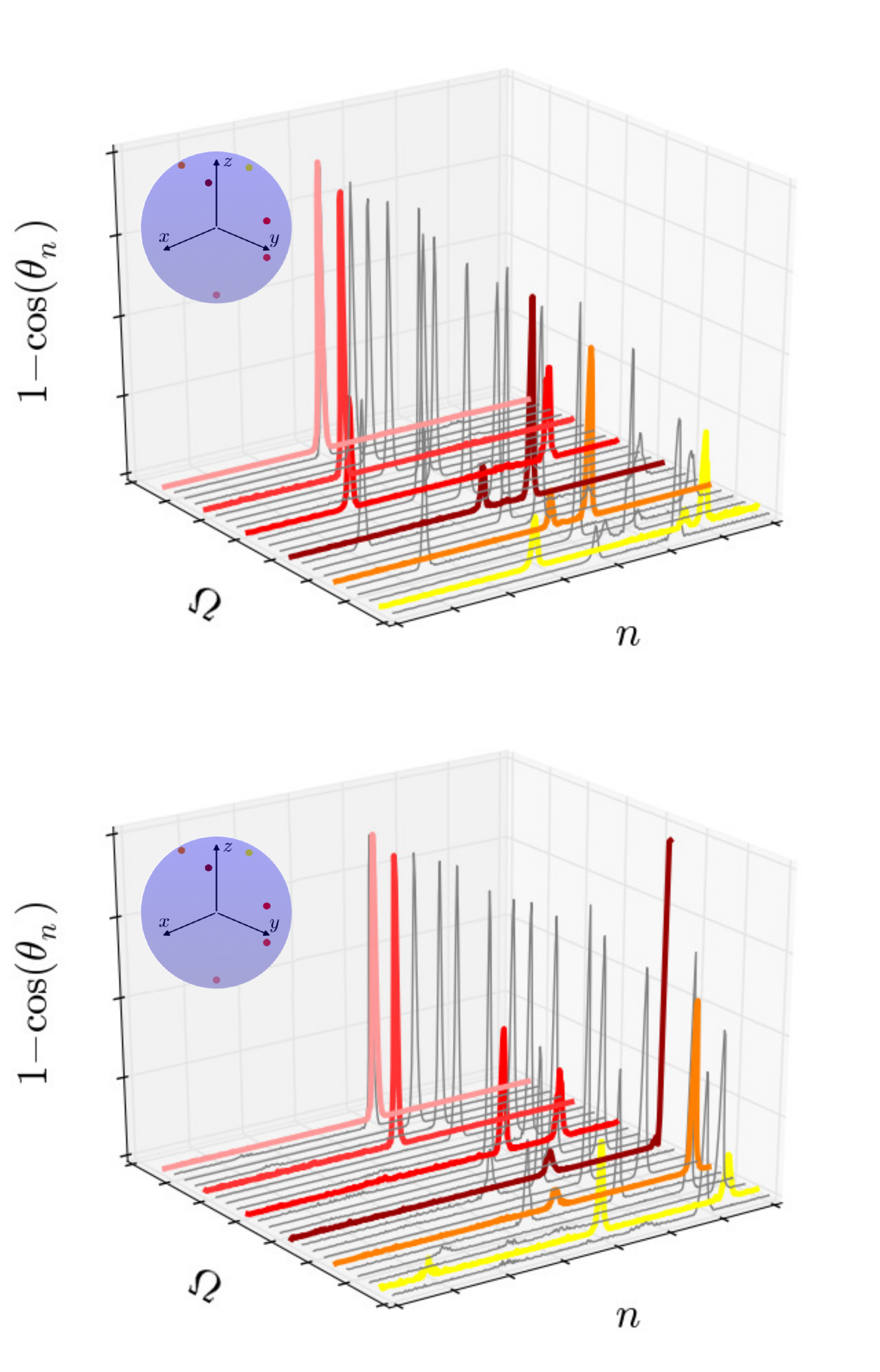}
	\caption{Same as in Fig.~\ref{f.3dconf}, for a starting soliton 
with $\lambda_\beta=2.5$  and integration times equal to 200 (upper 
panel) and 400 (lower panel). }
	\label{f.3dconf_2}
\end{figure}
Let us now concentrate upon the overlaps $\bbraket{\Omega_l(t, \Omega')}{\Omega_l(t, \Omega)}$ entering the 
above expression: if, for given $l$ and $t$, $\Omega_l(t,\Omega)$, 
only weakly depends on the initial 
value $\Omega$, the corresponding overlap is equal to one, the index $l$ disappears from 
Eq.~\eqref{e.t2_q1q2S2state}, and the spin $\bS_l$ effectively exits the dynamical scene.
If this is the case for all but a small number of adjacent spins of the chain, the 
entanglement originally generated by the interaction of $\qA$ with $\SA$ is not spread along 
the whole chain, but remains confined to the portion made of the above adjacent spins, whose configurations 
substantially depend on the initial value $\Omega$. 

In fact, this is precisely what happens in our setting when the initial 
configuration of the chain corresponds to a Heisenberg soliton whose 
width is larger than the chain spacing, as seen by comparing
Figs.~\ref{f.3dconf} and~\ref{f.3dconf_2}: 
while for $\lambda_\beta=2.5$ (Fig.~\ref{f.3dconf_2}) different $\Omega$
generate quite diverse configurations $\Omega_n(t,\Omega)$, if 
$\lambda_\beta=10$ (Fig.~\ref{f.3dconf}) the dependence of 
$\Omega_n(t,\Omega)$ on $\Omega$ is weaker and localized in a limited region of $\Gamma$. 
In this latter case, the soliton moves forward with a slightly modified shape and 
hauls the deformation of $\Omega_{\nA}(t_0)$ imposed while it traveled 
through site $\nA$. 
We can hence expect that, during the second stage of our scheme, the 
soliton behave as a carrier that keeps the entanglement localized while 
traveling along the chain.

\begin{figure}
	\includegraphics[width=0.45\textwidth]{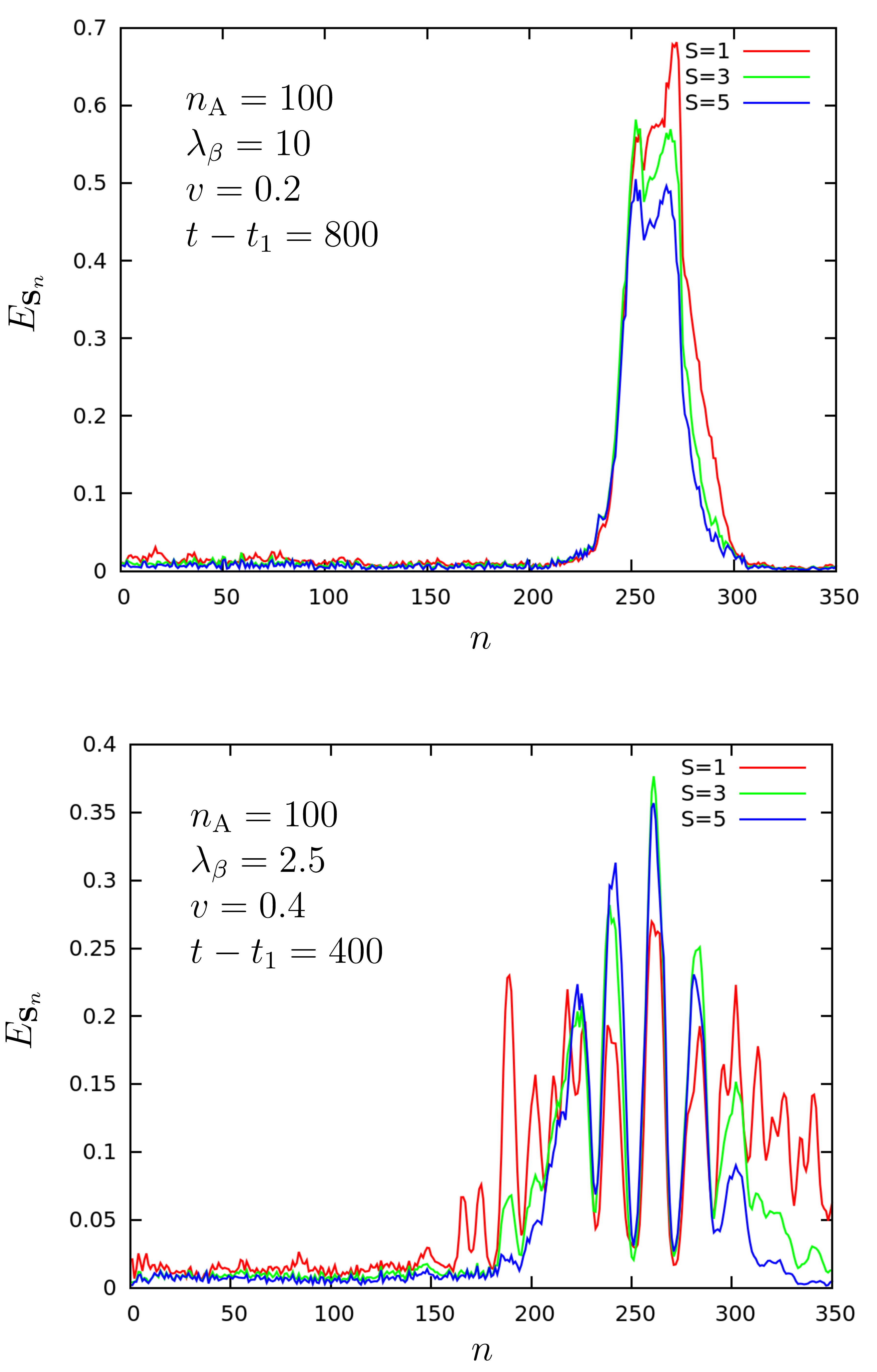}
	\caption
{$E_{{\bS_n}}(t)$ as a function of $n$ at
$t-t_1=800$ with a starting soliton of width 
$\lambda_\beta=10$ (upper panel) and at 
$t-t_1=400$ with a starting soliton of width 
$\lambda_\beta=2.5$ 
(lower panel). Curves for different values of $S$, as indicated.}
	\label{f.E2vsn}
\end{figure}

Numerical results do confirm this picture, as seen in 
Fig.~\ref{f.E2vsn}, where snapshots of $E_{{\bS_n}}$ are reported 
as a function of $n$. In the first panel of Fig.~\ref{f.E2vsn} a bump is 
clearly visible, centered at about $n=n_A+v(t-t_1)$, with $v$ 
the soliton velocity, which means that only the spins around the 
soliton are significantly entangled with the rest of the system. The 
different curves report the same 
quantity for different values of $S$: the shape is almost unchanged, but 
the values monotonically decrease with increasing $S$, according to the 
fact that in the limit $S\to\infty$ the entanglement disappears as the 
spins become completely classical.

The most favorable condition to establish entanglement between $\qA$ and $\qB$ is therefore achieved by choosing the time $t_2$ when the soliton crosses $\nB$: the superposition of the evolved configurations obtained from different deformations $\Omega$, Eq.~\eqref{e.q1chain_t2state}, is indeed expected to concentrate at such time around $\nB$ the entanglement collected at time $t_1$ in $\nA$. 

\section{Third stage: evolution of $(\qB,\SB)$}
\label{s.3rd}

During the third stage,  
$\qA$ is only affected by a uniform field, 
$\Gamma\setminus\SB$ does not evolve, and 
$\qB$ interacts with $\SB$ via the coupling $\gB(t)=g$.
 Apart 
from the different initial state, this stage is analogous to the first one with $\qA\leftrightarrow\qB$:
in fact, the propagator for $t\in[t_2,t_3]$ is
\begin{equation}
 \cU^{(3)}(t)= \cU_{\qA}(t) \otimes \mathds{1}_{\Gamma\setminus\SB} 
     \otimes \cU_{\qB,\SB}(t)\,,
\label{e.3rd_evo}
\end{equation}
to be compared with Eq.~\eqref{e.1st_evo_full}.
As we are interested in the entanglement between $\qA$ and $\qB$, we now have to 
determine the two-qubit density operator $\rho_{\qA\qB}(t)$.
Performing the partial trace of the projector \eqref{e.t2_sys_state}  
upon $\Gamma\setminus{\SB}$ at $t=t_2$ we obtain the initial state for the third stage, i.e.
\begin{eqnarray}
 \rho_{\qA{,}\SB{,}\qB}(t_2)&=&\Tr_{\Gamma\setminus{\SB}} \bket{\Psi(t_2)}\bbra{\Psi(t_2)}
\cr
&=&\rho_{\qA{,}\SB}(t_2) \otimes \bket{B(t_2)}\bbra{B(t_2)}~\label{e.ini_3rd},
\end{eqnarray}
with
\begin{eqnarray}
&& \rho_{\qA{,}\SB}(t_2) = {\cal A}^2 \sum_{\sigma\sigma'} \int d\Omega d\Omega'\,
     f_{\sigma\sigma'}^{\Omega\Omega'} \,\bket{\sigma (t_2)}\bbra{\sigma' (t_2)}
\notag\\
&& ~~~~~
 \times \bigg[ \prod_{n \neq \nB}
        \bbraket{\Omega_n(t_2, \Omega')}{\Omega_n(t_2, \Omega)} \bigg]
\notag\\
&& ~~~~~
    \otimes \ket{\Omega_{\rm{B}}(t_2,\Omega)}\bra{\Omega_{\rm{B}}(t_2,\Omega')}~,
\end{eqnarray}
where $\Omega_{\rm{B}} \equiv \Omega_{\nB}$.
Notice that, having traced out all the spins of $\Gamma$ but $\SB$, for $t > t_2$ we deal with the Hilbert 
space of $\qA,\qB$ and $\SB$ only, which has dimension $4(2S{+}1)$ no matter the distance between the qubits, i.e. 
the length of the portion of chain between $\SA$ and $\SB$. 
The propagator for $(\qA,\SB,\qB)$ is $\exp\{-i\cH_{\qA{,}\SB{,}\qB}\,t\}$,
with
\begin{equation}
 \cH_{\qA{,}\SB{,}\qB} = \hA \hsigA^z
 + g\,\hSB{\cdot}\,\hbsigB + \hB \hsigB^z~,
 \label{e.3rd_ham}
\end{equation}
that can be diagonalized numerically \cite{note3}; the generic 
element of the density matrix of $(\qA,\SB)$ can be written as
\begin{eqnarray}
&&\big[\rho_{{\qA{,}\SB}}(t_2) \, \big]_{mm'}^{\sigma\sigma'}
 = {\cal A}^2 e^{-ih(t_2-t_1)(\sigma{-}\sigma')} \int d\Omega\,d\Omega'\,
 f_{\sigma\sigma'}^{\Omega\Omega'}
\notag \\  
&&~\times~
 \bigg[\prod_{n \neq \nB} \bbraket{\Omega_n(t_{2},\Omega')}{\Omega_n(t_{2},\Omega)} \bigg]
 {\cal D}_{mm'}^{\Omega\Omega'}(t_2) ~,
\label{e.q1s2_rhomrep}
\end{eqnarray}
with 
\begin{equation}
 {\cal D}_{mm'}^{\Omega\Omega'}(t_2) = 
 \bbraket{m}{\Omega_{\rm{B}}(t_{2},\Omega)} \bbraket{\Omega_{\rm{B}}(t_{2},\Omega')}{m'}\,.
\end{equation} 

Given Eqs.~\eqref{e.ini_3rd}--\eqref{e.q1s2_rhomrep}, making use of the relations~\eqref{e.CS_mO} 
and~\eqref{e.CSoverlap} one can numerically compute $\rho_{{\qA{,}\SB{,}\qB}}(t_2)$ and its evolved state with $t$ any time larger than $t_2$. 
Although $\rho_{{\qA{,}\SB{,}\qB}}(t)$ is in general a non-separable state of $\qB$ and $(\qA,\SB)$, this does not necessarily mean that
$\qA$ and $\qB$ are entangled. In order to settle this, one has to trace out $\SB$, yielding the two-qubit density operator
\begin{equation}
 \rho_{{\qA{,}\qB}}(t) = \Tr_{\SB} \left[ \rho_{{\qA{,}\SB{,}\qB}}(t) \right] ~,
\label{e.rhoq1q2}
\end{equation}
and evaluate the {\it concurrence}~\cite{Wootters1998} between $\qA$ and $\qB$, defined, for any two-qubit density 
operator $\rho$, as
\begin{equation}
 {\cal C}(\rho) \equiv \max (0,\mu_1-\mu_2-\mu_3-\mu_4)\,,
\label{e.concurrence}
\end{equation}
where $\{\mu_1^2,\,\mu_2^2,\,\mu_3^2,\,\mu_4^2\}$ are the eigenvalues 
(in decreasing order) of the Hermitian operator
$\sqrt{\rho}\,\tilde\rho\,\sqrt{\rho}$ with 
$\tilde\rho=(\sigma^y\otimes\sigma^y)\rho^*(\sigma^y\otimes\sigma^y)$.

\begin{figure}
\includegraphics[width=0.45\textwidth]{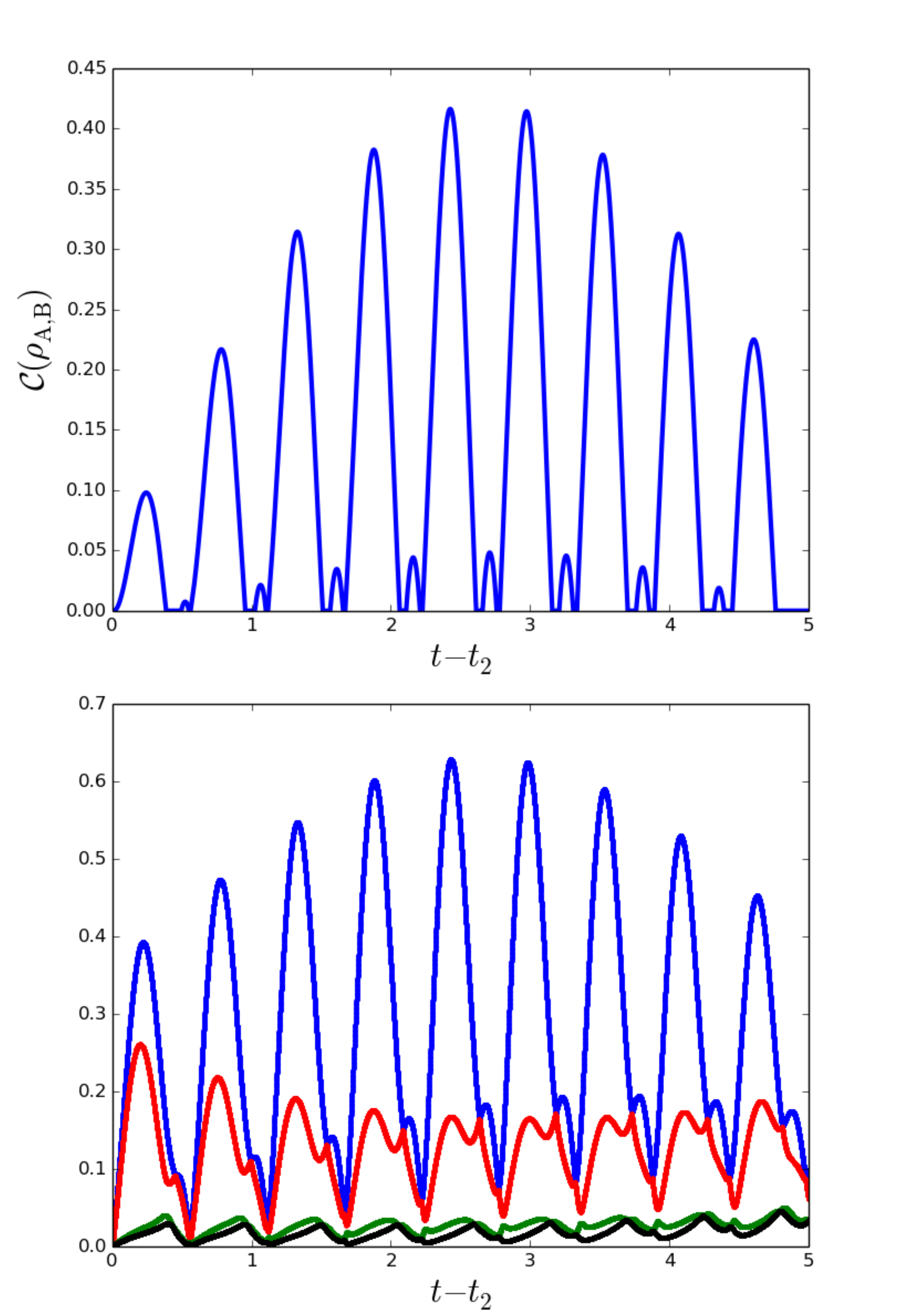}
 \caption{Upper panel: ${\cal C}[\rho_{\qA,\qB}(t-t_2)]$ for $g=1$, 
$\hA=\hB=0.25$, $S=5$, and a starting soliton with $\lambda_\beta=10$.
Lower panel: eigenvalues $\{\mu_1,\mu_2,\mu_3,\mu_4\}$ for the same 
parameters values as above. Cusps in the upper panel originates from the 
eigenvalues crossings seen in the lower panel.}
\label{f.concurrence}
\end{figure}

An example of the concurrence ${\cal C}[\rho_{\qA,\qB}(t)]$ for $t>t_2$ is shown in Fig.~\ref{f.concurrence}: this is obtained starting from the initial state \eqref{e.sys_iniconf}
with  $\ket{A}=\ket{B}=\ket1$ (i.e., the eigenstate of $\sigma^z$ with eigenvalue +1) and $\{\Omega_n (t_0) \}$ the 
configuration corresponding to a Heisenberg soliton of width $\lambda_\beta\,{=}\,10$, centered at $\SA$ at 
$t=t_0$. The figure shows finite time-intervals during which ${\cal C}(\rho_{\qA,\qB})$ 
is significantly different from zero, implying that there exist values of $t_3$ when to quench the 
($\qB$,$\SB$) interaction so as to leave the qubit-pair in a stationary and entangled state.
Notice that the periodic exchange of entanglement between $\SB$ and $\qB$ is a consequence of the dynamics 
ruled by the Hamiltonian~\eqref{e.3rd_ham}. 

\begin{figure}
\includegraphics[width=0.45\textwidth]{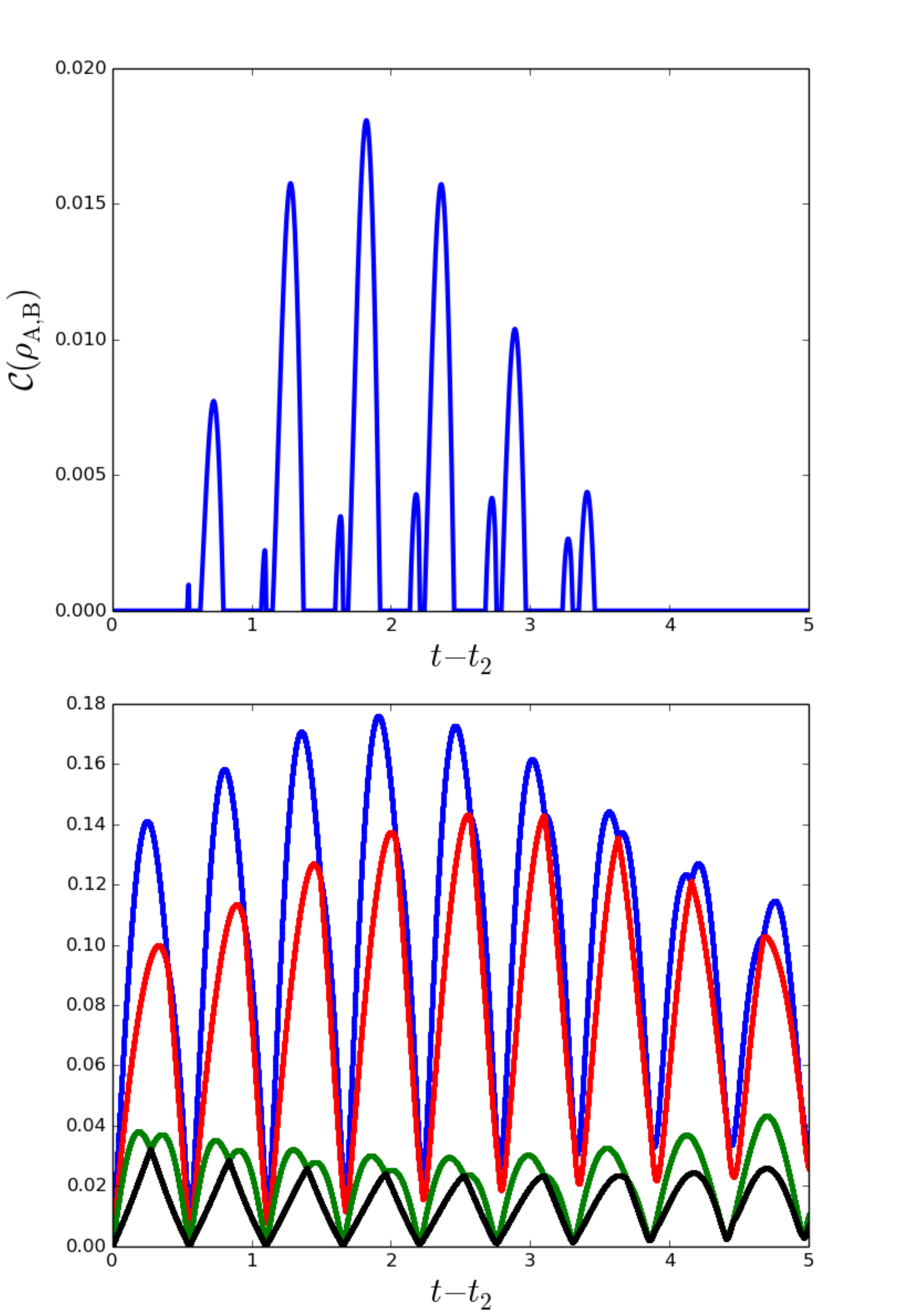}
\caption{Same as in Fig.~\ref{f.concurrence} for a starting soliton with
$\lambda_\beta=2.5$.}

\label{f.conc2}
\end{figure}

It is further observed that choosing different values for the parameters (the spin value $S$, the couplings 
$\gA$, $\gB$, the local fields $\hA$, $\hB$, etc.), or a different initial configuration $\{\Omega_n (t_0) \}$ 
(i.e., a different soliton) for the initial state of $\Gamma$, does not qualitatively affect the 
numerical results for ${\cal C}[\rho_{\qA,\qB}(t)]$, that keeps displaying the oscillatory behavior observed in 
Fig.~\ref{f.concurrence} although with different frequency and peaks of different heights.
In fact, these heights are found significantly different from zero if {\it i)} $t_2\sim t_1 + (\nB 
-\nA)/v$, implying that $\SB$ is amongst the spins which are correlated with both the rest of the 
chain and the qubit $\qA$, and {\it ii)} the initial state of $\Gamma$ is
such that the superposition~\eqref{e.q1chain_t2state} allows 
for the entanglement to be localized on a small number of spins rather than on a large portion of the chain. For 
instance, referring to Fig.~\ref{f.E2vsn}, higher values of the concurrence are found when the 
situation shown in the upper panel occurs, as seen by comparing Figs.~\ref{f.concurrence} and~\ref{f.conc2}.
Overall, choosing $t_3$ such that ${\cal{C}}\big[\rho_{{\qA{,}\qB}}(t_3)\big]\neq0$, the evolution of the 
proposed model takes a separable state of $\qA$ and $\qB$ into an entangled one for the pair, 
thus behaving as an entangling gate.

\section{Conclusions}
\label{s.conclusions}

The results presented in the previous sections show that 
a large-$S$ spin chain can be employed to generate entanglement 
between two distant qubits $\qA$ and $\qB$. The spin chain initially 
is in a classical-like state, corresponding to a running Heisenberg 
soliton passing by $\qA$. In a first stage $\qA$ interacts with the 
chain spin $\SA$, dynamically establishing quantum correlations 
which, in a second stage, the moving soliton can efficiently carry 
with to the location of the chain spin $\SB$, which in turn interacts 
with $\qB$ in such a way that finally the system of the two qubits is 
in an entangled state.

The one-to-one mapping between the classical spin chain 
configurations and the tensor product of single-spin coherent states, 
allowed us to approximate the quantum evolution of the chain. 
However, in order to obtain the final quantum state, several 
classical-like evolutions must be superposed, as after the first 
stage $\SA$ is no more in a definite coherent state: such a 
simultaneous existence of `parallel classical histories' explains why 
a classical-like description of the chain dynamics can account for 
quantum correlation transfer.
 
The explicit calculations have been made feasible by the introduction of 
simplifying assumptions.
The first one concerns the time dependence of the qubit-chain 
interactions, which implies the ability to somehow switch on and off 
the interaction in a very short time: although this is a typical 
approximation in theoretical schemes, it is not always clear how to 
implement it in diverse realizations, especially for solid-state 
devices. The onset of the entangling dynamics between $\qA$ and 
$\SA$, and later on between $\qB$ and $\SB$, in the terms described 
in Secs.~\ref{s.1st} and~\ref{s.3rd}, can be thought to be embedded 
in the original model: in fact, before the soliton arrival all spins 
and the qubits are in the {\em up} state, so the interactions act 
trivially, giving an overall phase factor; only when the incoming 
soliton modifies the state of $\SA$, a non trivial dynamics of the 
($\qA$,$\SA$) subsystem is induced; in a similar way, the relevant 
dynamics of the subsystem($\qB$,$\SB$) only starts when the partially 
deformed soliton, reaches $\SB$. This would effectively be tantamount 
to switching on the couplings between the qubits and the 
chain, although not abruptly as in Eq.~\eqref{e.intconst}, and it is not 
to be expected to yield dramatic changes in the qualitative behavior. 
A suitable mechanism for finally quenching the interactions can also 
be imagined, as for instance that proposed in~\cite{HeintzeEtal13}.

A further simplification was to assume the chain to be `frozen' 
during the evolutions of the pairs $(\qA,\SA)$ and $(\qB,\SB)$ (first 
and third dynamical stage), i.e., that the typical time-scale of the 
qubit-spin interaction, $(gS)^{-1}$, be much smaller than that of the 
chain dynamics associated to the propagating soliton, given by 
$\tau_{\beta}=(JS\,h\,\sin\!2\beta)^{-1}$ (see 
Appendix~\ref{a.Hsolitons}), namely,
\begin{equation}
  {J\over g}{h\sin 2 \beta} \ll 1 ~.
\label{e.tscaleratio}
\end{equation}
The above relation can be satisfied both if $g\gg J$, i.e., the chain 
coupling is much weaker than the qubit-spin coupling, or if $h\simeq 
\mu_B H/ JS\ll 1$, i.e., the intensity $H$ of the uniform field 
applied to the chain is weak compared with the chain coupling. This 
second requirement is usually met if the spin chain is thought to be 
some solid-state system, as typically exchange energies are much 
larger than Zeeman energies.

In virtue of the described results we conclude that, by choosing 
suitable values of the tunable parameters and the initial state, a large-$S$ 
spin chain can realize a two-qubit entangling gate. The carriers of quantum 
correlations, i.e., solitons, are known to be robust against noise 
and external disturbances, and make the hybrid scheme we have proposed a 
promising alternative to the most commonly studied purely quantum buses.

\begin{acknowledgments}
We acknowledge financial support from the University of Florence in the framework of the University Strategic Project Program 2015 (project BRS00215).
This work was performed in the framework of the Convenzione operativa between the Institute for Complex
Systems of the Italian National Research Council, and the Physics and Astronomy Department of the University
of Florence. 
\end{acknowledgments}

\appendix

\section{Heisenberg solitons}
\label{a.Hsolitons}

Let us consider a classical spin-$S$ Heisenberg chain, i.e., a 1D array of (spin-)vectors $\bm{S}_n\equiv{S}\,\bm{s}_n$, whose magnitude $S$ has the dimension of an action. The unit vectors $\bm{s}_n$ are naturally parametrized by polar coordinates, $\bm{s}_n\equiv(\sin\theta_n\cos\varphi_n,\sin\theta_n\sin\varphi_n, \cos\theta_n)$, with $\varphi_n$ and $\cos\theta_n$ canonically conjugated variables, $\{\varphi_n,\cos\theta_l\}=S^{-1}\,\delta_{nl}$. Its Hamiltonian is the classical analogue of Eq. (\ref{e.hchain}):
\begin{equation}
{\cal H}_{\rm cl}= -JS^2 {\sum}_n \bm{s}_n{\cdot}\bm{s}_{n+1}
  -\gamma S\bm{H}{\cdot}{\sum}_n \bm{s}_n~,
\label{e.Hdiscr}
\end{equation}
and the corresponding EoM for the unit vectors 
$\bm{s}_n$ are 
\begin{equation}
\partial_t\bm{s}_n=JS\,\bm{s}_n\times(\bm{s}_{n+1}{+}\bm{s}_{n-1}+\bm{h})~,
\label{e.classical_EoM}
\end{equation}
where $JS$ sets the frequency scale and $\bm{h}\equiv\gamma\bm{H}/(JS)$ is the dimensionless Zeeman field.

\begin{figure}
	\centerline{
	\includegraphics[width=0.45\textwidth]{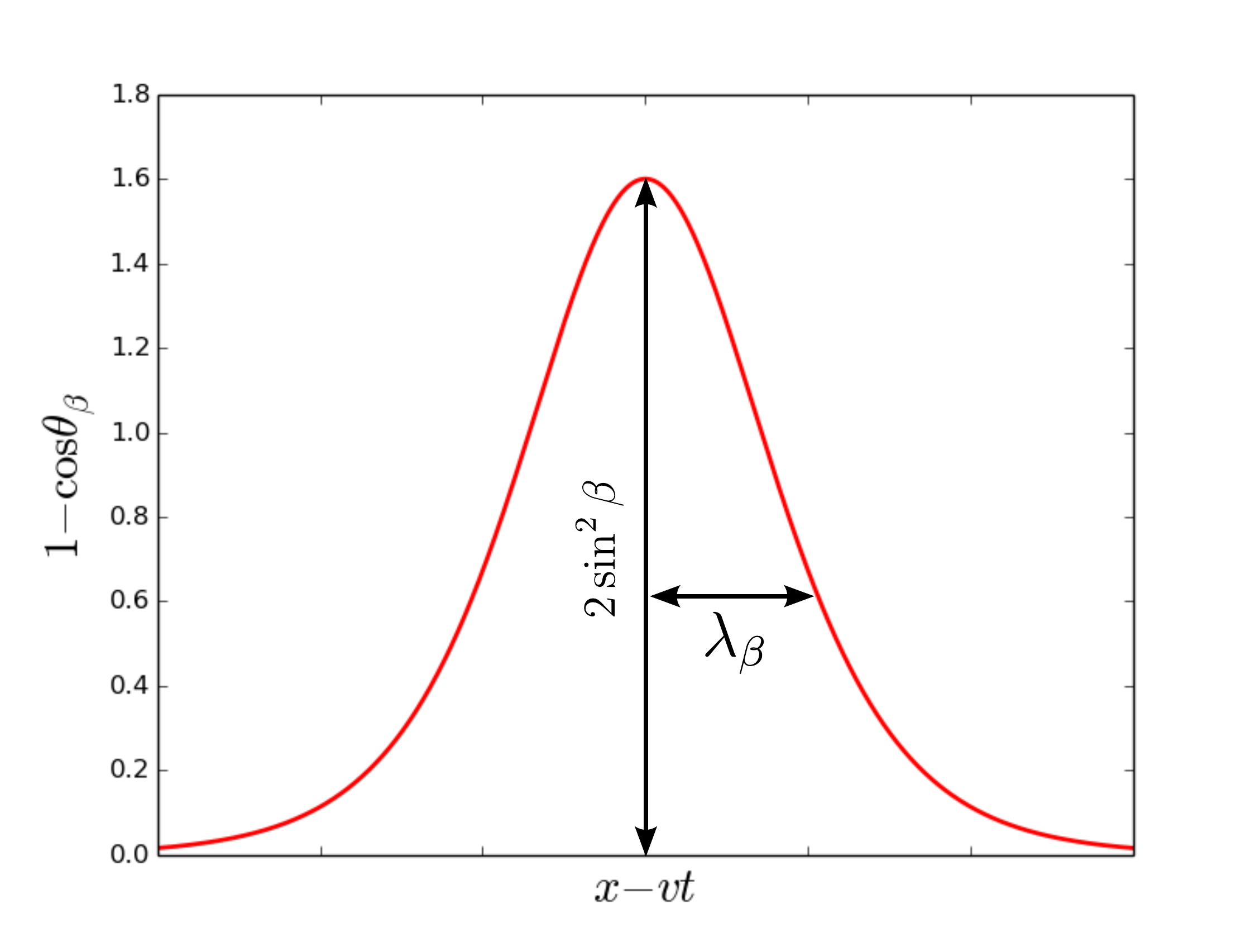}
	}
	\caption{TW soliton:~ $1-\cos\theta_\beta(\xi)$ for $\tan\!\beta\,{=}\,2$.}
	\label{f.soliton}
\end{figure}
As shown by Tjon and Wright~\cite{TjonW1977} (TW), the Heisenberg chain EoM  have, in the continuum approximation (lattice spacing $d\,{\to}\,0$), an analytical `one-soliton' solution of the form:
\begin{equation}
\left\{
\begin{aligned}
 \theta_{\beta} &= 2\sin^{-1}(\sin\!\beta\,\sech\xi) ~,
\\
\varphi_{\beta} &= \varphi_0 + \cot\!\beta\,\xi
+\tan^{-1}(\tan\!\beta\,\tanh\xi) ~,
\end{aligned}
\right.
\label{e.twsol}
\end{equation}
where
$\xi\equiv\,(x{-}vt)/\lambda_{\beta}$ and $x\,{=}\,nd$ is the `continuum' coordinate. The soliton \emph{amplitude}, characterized by the dimensionless parameter $\beta\in(0,\pi/2)$ ($\theta\,{\le}\,2\beta$) is related with the soliton velocity $v$ by 
$\cos\!\beta=v/({2dJS\sqrt{h}})$, %\label{e.solvel}
and determines the soliton \emph{length}~
$\lambda_{\beta}=d/({\sqrt{h}\sin\!\beta})$~, %\label{e.sollength}
\emph{time scale}~
$\tau_{\beta}=(JS\,h\,\sin\!2\beta)^{-1}$~, %\label{e.soltime}
and {\em energy}~
$\varepsilon_{\beta}=8JS^2\sqrt{h}\,\sin\!\beta$~. %\label{e.solen}
Fig.~\ref{f.soliton} reports a typical TW soliton, and it is useful to note that solitons with larger amplitude $\beta$ have larger energy ($\sim\sin\!\beta$), are narrower $(\sim1/\sin\!\beta)$ and slower ($\sim\cos\!\beta$). 
Although there are no known analytic soliton solutions of the discrete model, the continuum approximation holds for configurations that vary slowly on the scale of the lattice spacing $d$, so that the solution~\eqref{e.twsol} approximately applies also to the chain model~\eqref{e.Hdiscr} provided that $\lambda_{\beta}\gg{d}$, i.e., $\sqrt{h}~\sin\!\beta\,{\ll}\,1$. This is generally true in real systems, whose typical exchange energies are of the order of tenths-hundreds of Kelvin degrees: as $\mu_{\rm{B}}=0.67$~K/Tesla, only very large fields could break the inequality.
Numerical investigations confirmed that soliton-like excitations can be injected in discrete spin chains and propagate along them without substantial distortion~\cite{CNVV2015a}.

\section{Spin coherent states}
\label{a.spinCS}

The states of a spin-$S$ particle are usually expanded on the basis of the $2S{+}1$ eigenvectors of the $z$-component of the spin operator, $\hat{S}^z\ket{m}=m\ket{m}$, with $m=-S,\,...,S$.
Given an arbitrary direction ${\bm s}$ in 3D space, i.e., a unit vector ${\bm s}\equiv(\sin\theta\cos\varphi,\sin\theta\sin\varphi,\cos\theta)$ defined by its spherical angles $\{\theta,\varphi\} \equiv \Omega$, the corresponding spin coherent state $\ket\Omega$ is defined as
\begin{equation}
\ket{\Omega} = \big(\cos\fr\theta{2}\big)^{2S}\,
              \exp \Big(\tan\fr\theta{2}\,e^{i\varphi}\,\hat{S}^- \Big) \ket S ~,
\label{e.coher_def}
\end{equation}
$\ket{S}$ being the eigenvector of $\hat{S}^z$ with maximal eigenvalue,  $m{=}S$. This state can also be written in the usual basis of eigenstates of $\hat{S}^z$:
\begin{equation}
 \ket{\Omega} = \sum_{m=-S}^{S} \braket{m}{\Omega} \ket m \,,
\label{e.CS_Om}
\end{equation}
the coefficients in this relation being the overlaps between the eigenvectors $\ket{m}$ and the coherent state $\ket\Omega$,
\begin{eqnarray}
 \braket{m}{\Omega} &=&  \big(\cos\fr\theta{2}\big)^{2S} \sqrt{\textstyle\frac{2S!}{(S{-}m)!(S{+}m)!}} 
\notag\\
&&~~~~~~~~\times~\big(\tan \fr\theta{2}\big)^{(S-m)} e^{i(S-m)\varphi} ~.
\label{e.CS_mO}
\end{eqnarray}

An important property of spin coherent states is that the expectation values of the spin-component operators are equal to the components of a classical vector of modulus $S$ oriented along ${\bm s}$, i.e.,
\begin{equation}
 \braket{\Omega| \hat\bS }{\Omega}
 = S(\sin\theta\cos\varphi,\sin\theta\sin\varphi,\cos\theta)=S\,{\bm s} ~.
\label{e.CSaveS}
\end{equation} 
Spin coherent states form a non-orthogonal and overcomplete set of states. Indeed,
\begin{equation}
 \braket{\Omega'}{\Omega}= \Big(\cos\fr\theta{2} \cos\fr{\theta'}{2}+ \sin\fr\theta{2} \sin\fr{\theta'}{2}\,e^{i(\varphi-\varphi')}\Big)^{2S} \,,
\label{e.CSoverlap}
\end{equation}
which implies 
\begin{equation}
 \big|\braket{\Omega'}{\Omega}\big|^2 
 = \Big( \fr{1+{\Omega}\cdot{\Omega}'}{2} \Big)^{2S} 
 = \Big( \cos\fr{\widehat{\Omega\Omega'}}{2} \Big)^{4S} ~,
 \label{e.CSoverlap2}
\end{equation}
i.e., the overlap modulus depends on the angle $\widehat{\Omega\Omega'}$ between the directions identified by $\Omega$ and $\Omega'$, respectively. The (over)completeness relation reads
\begin{equation}
 (2S{+}1)\int\frac{d\Omega}{4\pi}\,\ket\Omega \bra\Omega = \mathds{1}_S ~,
\label{e.CScompl}
\end{equation}
where $d\Omega={d\cos\theta}~{d\varphi}$.

From Eq.~(\ref{e.CSoverlap2}) we see that $\big|\braket{\Omega'}{\Omega}\big|^2\propto\delta(\Omega-\Omega')$ in the 
limit $S\to\infty$, and reminding Eq.~(\ref{e.CSaveS}) and the description of the classical Heisenberg chain given in Appendix~\ref{a.Hsolitons}, it clearly appears that the spin coherent states are the tool of choice to properly address the classical limit of spin systems.


\begin{thebibliography}{00}
\bibitem{note1}
The validity of this approximation in a realistic setup sets constraints on the ratios of the interaction constants $\gA$, $\gB$ and the spin-chain time scale; nevertheless, this extreme simplification is aimed at demonstrating the possibility of entanglement transfer by a quasi-classical channel.
\bibitem{note2} The Von Neumann entropy $E_{{\qA,\SA}}(\bket{\psi_{\qA,\SA}(t)})$ of $\qA$ with $\SA$ is given by
$E_{{\qA,\SA}}(\bket{\psi_{\qA,\SA}(t)})=-\Tr_{\qA}\rho_{\qA} \log_2 \rho_{\qA}$ with $\rho_{\qA}=\Tr_{\SA} \bket{\psi_{\qA,\SA}(t)}\bbra{\psi_{\qA,\SA}(t)}$. Since the rest of the system is factorized at this stage of the evolution, $E_{{\qA,\SA}}$ correctly quantifies the entanglement between $\qA$ and $\SA$.
\bibitem{note3}
The evolution of the initial state is evaluated exactly by the numerical diagonalization of the Hamiltonian~\eqref{e.ham_q1s1}; numerical calculations are performed using the basis of the $\hat S^z$ and $\hat\sigma^z$ eigenvectors:\\
$\big\{ {\ket{m}\ket{\sigma}{\equiv}\ket{m}{\otimes}\ket\sigma} \,\big|\,
        {m{\in}\{-S,\,...,\,S{-}1,\,S\},\,\sigma{\in}\{1,-1\}} \big\}$\,.
\end{thebibliography}
\end{document}